\documentclass{cernrep}
\usepackage{lineno}
\begin{document}
\title{Search for Ultra-High Energy Photons with the Pierre Auger Observatory}
\author{P. Homola\thanks{Institute of Nuclear Physics Polish Academy of Sciences, Krak\'ow, Poland} 
for the Pierre Auger Collaboration\thanks{Av. San Mart\'in Norte 304 (5613) Malarg\"{u}e, Prov. de Mendoza, Argentina}
\thanks{Full author list: http://www.auger.org/archive/authors\_2017\_08.html}}


\begin{abstract}
One of key scientific objectives of the Pierre Auger Observatory is 
the search for ultra-high energy photons. Such photons could originate 
either in the interactions of energetic cosmic-ray nuclei with the cosmic 
microwave background (so-called cosmogenic photons) or in the exotic scenarios, 
e.g. those assuming a production and decay of some hypothetical super-massive 
particles. The latter category of models would imply relatively large fluxes 
of photons with ultra-high energies at Earth, while the former, 
involving interactions of cosmic-ray nuclei with the microwave background - 
just the contrary: very small fractions. The investigations on the data collected 
so far in the Pierre Auger Observatory led to placing very stringent limits 
to ultra-high energy photon fluxes: below the predictions of the most of 
the exotic models and nearing the predicted fluxes of the cosmogenic photons. 
In this paper the status of these investigations and perspectives for further studies
are summarized.
\end{abstract}


\maketitle

\section{Introduction}
\label{sec:intro}
So far, photon structure is most efficiently studied with the accelerator instruments, where
available energies reach TeV. There is, 
however, some potential in exploring the cosmogenic photons as well. Despite the cosmic photon 
fluxes being extremely low in comparison to accelerator data, different physics mechanisms are involved at the production
sites and during the subsequent propagation to the Earth. This gives prospects for a complementary
study on photons. Also the energies of cosmogenic photons can be higher than in accelerators. Gamma astronomers 
report on power-law spectra of $\gamma$-rays extending without a cut-off or a spectral break to tens of TeV
\cite{hess-pevatrons-nature2016} and plan building instruments capable of detecting photons of energies up to 300~TeV
\cite{cta}. But these are not the largest photon energies expected on Earth. Within the research concerning 
ultra-high energy cosmic rays (UHECR), i.e. particles with energies exceeding 10$^{18}$ eV,
all the prominent scenarios predict photon fluxes reaching the Earth. 
One distinguishes two major classes of UHECR models: ``bottom-up'', 
based on the acceleration and subsequent interaction of nuclei; 
and ``top-down'', based on a 
decay or annihilation of hypothetical supermassive particles (see \cite{Bhattacharjee:1998qc} for a review). 
The two classes differ significantly 
in the predicted fractions of photons among UHECR: < 1\% the former and even up to 
c.a.~50\% the latter. Hence, determining the UHECR mass composition, including identification of
photons or setting upper limits on their fluxes, is an effort towards distinguishing between the two major classes
which should give a hint on photon production and properties at the highest energies. 
Moreover, it is worth emphasizing that any result
on UHE photons, including non-observation, is meaningful for the foundations of physics at the highest energies,
allowing constraints on e.g. Lorentz invariance violation (LIV) \cite{Galaverni:2007tq}, 
QED nonlinearities \cite{maccione08}, space-time structure \cite{maccione-liv-spacetime-foam-2010} 
or the already mentioned ``top-down'' scenarios.

The searches performed so far have not confirmed the existence of UHE photons, resulting in setting upper limits
to photon fluxes and fractions 
\cite{Aglietta:2007yx,Abraham:2009qb,auger-photons-bleve2015,ta-photons-zayyad-2013,ta-photons-rubtsov2015}. 
In this paper we summarize the recent advances in UHE photon search performed
based on the data collected by the largest cosmic-ray instrument: the Pierre Auger 
Observatory~\cite{auger-2015,auger-upgrade} resulting
in setting the most stringent upper limits. The summary is based
on Refs.~\cite{auger-photon-pointsources-2014},\cite{auger-diffuse-photon-2017}, and 
\cite{auger-targeted-photon-2017}. We also briefly sketch an outlook on the further possible efforts, 
complementary to the current searches, based on the cascade approach.  

\section{Photon signatures}
\label{sec:signatures}

Investigations on cosmic rays of energies above 10$^{15}$~eV can be done only indirectly, through the interpretation of
properties of extensive air showers (EAS) induced in the atmosphere by primary cosmic particles. The interpretation is made
based on interaction models which incorporate cross section data from accelerators within 
the available energy range and assume extrapolations beyond this range. Obviously, the higher the primary energy under
investigation, the higher the risk of being mistaken with the extrapolations being used to interpret the data. Another 
disadvantage and potential uncertainty is the progressing degeneracy of the information about primary particles 
and their first interactions in the atmosphere with the rise of subsequent generations of secondary particles. 
Thus, if there is some mistreatment in the models at the very beginning of an air shower creation and propagation,
it might significantly affect the final outcome of the analysis. Keeping in mind all the theoretical assumptions implying
the existence of uncertainties, an effort is being made to identify the ultra-high energy primary particles
that initiate the largest air showers observed. The effort is based on using simulations to predict air
shower properties, characteristic for primaries of different types and confronting 
these predictions with observations. 

EAS initiated by UHE photons should posses two 
independent properties:
significantly delayed development and reduced muon content \cite{Risse:2007sd}. 
The former signature is based on the assumption that at the highest energies the pair production formation length
gets elongated so much that the destructive interference of the fields associated with the atoms and particles 
nearby the primary
suppresses the pair production process in the upper atomosphere, thus delaying the first interaction 
and the consequent air shower development. This is so-called LPM effect \cite{Landau:1953um,Landau:1953gr,Migdal:1956tc}, 
a standard in air 
shower modeling \cite{corsika}.
Here it is important to note that the LPM suppression is sensitive to the possible LIV effects
including both increase and reduction of the pair production formation length\cite{Vankov:2002gt}. While the 
former would strengthen the UHE photon discrimination power based on delayed air shower development, the latter would 
do the opposite: reduce or even invert the LPM effect, and hence make photon-induced air showers develop more similarly
to those initiated by nuclei. The present state-of the art photon identification methods involving the expectation of a 
delayed development of an air shower induced by a photon assume the standard LPM effect, without any LIV effects.

The other expected property of photon-induced air showers, the low muon content N$_{\mu}$, is also founded on 
conventional physics assumptions concerning the photonuclear cross section $\sigma_{\gamma-air}$. Air shower 
muons are thought to originate mostly from the charged pion decays, and charged pions originate from hadronic interactions.
Therefore the observed N$_{\mu}$ should correspond to the size of a hadronic component of an air shower. 
If the interaction initiating an air shower is not hadronic, one would consequently expect that the
hadronic component is initiated only by one of the secondary particles, thus its size and the corresponding
N$_{\mu}$ are smaller comparing to an air shower induced by a hadronic interaction.
According to the standard extrapolations for a photon of energy $E_{\gamma}=10^{19}$~eV, 
$\sigma_{\gamma-air}$ should be ca.~30 times smaller than the electron pair production cross section, although some
existing models allow $\sigma_{\gamma-air}$ to be only ca. 3 times smaller at the highest energies \cite{Risse:2007sd}.
The energies of the secondary and virtual photons inside air showers are naturally lower than the primary
energy and thus the $\sigma_{\gamma-air}$ uncertainty is correspondingly lower. Since in any case one expects that
primary photons at the highest energies would pair produce more readily than interact with air molecules, the 
hadronic component and N$_{\mu}$ in the corresponding air showers will be in general smaller than in case of hadron-induced
showers. The data collected so far by the Pierre Auger Observatory do not reveal muon-poor showers, just the opposite 
-- the muon
content seems to exceed the simulated values at the highest energies \cite{auger-muon-excess16}, 
which might point to a yet unconsidered physical 
source of uncertainty related to N$_{\mu}$.

\section{Diffuse photon flux and hybrid UHE photon limits with multivariate analysis}

\begin{figure}[ht]
\begin{center}
\includegraphics[width=\linewidth]{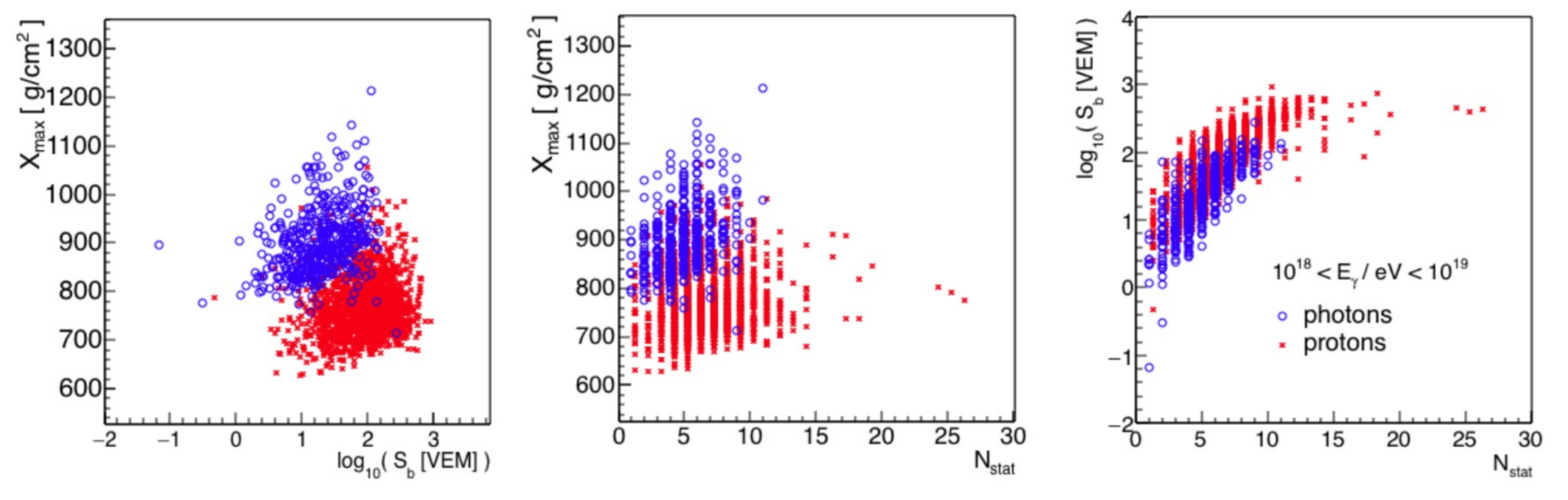}
\caption{Photon-proton discrimination power of the shower observables used for the UHE photon search in the hybrid 
detector of the Pierre Auger Observatory.\cite{auger-diffuse-photon-2017}}
\label{fig:signatures}
\end{center}
\end{figure}

\begin{figure}[ht]
\begin{center}
\includegraphics[width=\linewidth]{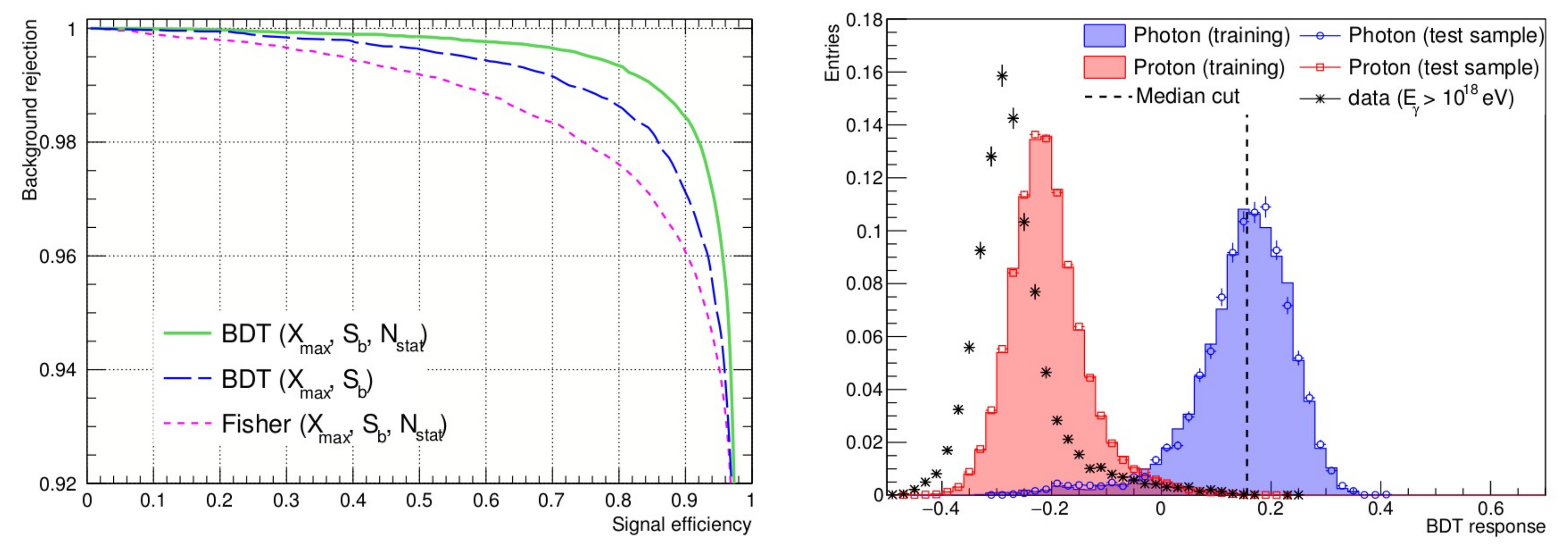}
\caption{The variants of the multivariate analysis method considered for the diffuse photon flux studies (left)
and the selection cut on the response of the optimum variable (right). \cite{auger-diffuse-photon-2017}}
\label{fig:signatures-bdt}
\end{center}
\end{figure}

\begin{figure}[ht]
\begin{center}
\includegraphics[width=0.6\linewidth]{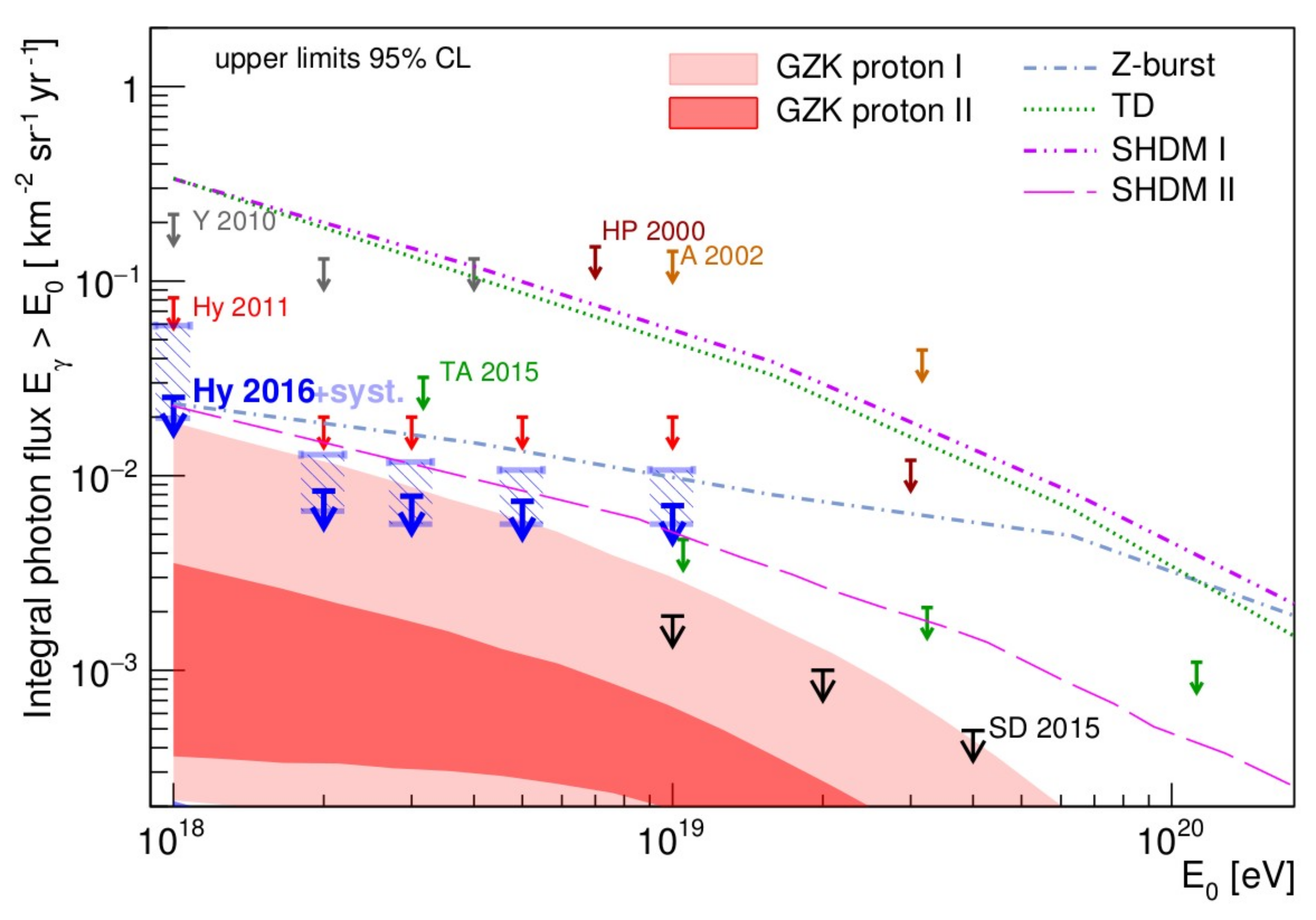}
\caption{The newest hybrid upper limits to the diffuse photon fluxes based on the 
Pierre Auger Observatory data (blue arrows) and the corresponding uncertainties
(dashed regions around the arrows) compared to the theoretical predictions (check Ref.~\cite{auger-diffuse-photon-2017} for 
references) and to the limits provided by other experiments.~\cite{auger-diffuse-photon-2017}}
\label{fig:limits}
\end{center}
\end{figure}

In this section we follow Ref.~\cite{auger-diffuse-photon-2017} to present the upper limits to 
diffuse UHE photon fluxes obtained using the hybrid detector of the Pierre Auger Observatory. The observables 
used for this analysis are tightly connected with the key photon signatures introduced in the previous section: 
$X_{max}$ [g/cm$^2$] -- the depth in the atmosphere where an EAS reaches maximum of its development, 
$N_{stat}$ -- the number of 
triggered stations in an effective EAS footprint composed of the triggered stations, and
$S_b = \sum_{i}^N S_i (R_i/R_0)^b$, 
where $S_i$ and $R_i$ are the signal and the distance from the shower axis of the $i$-th station,
$R_0$~=~1000~m is a reference distance and b = 4 is a constant optimized to have the best
separation power between photon and nuclear primaries in the energy region above 10$^{18}$~eV. $S_i$ are measured 
in units of VEM (Vertical
Equivalent Muon, i.e. the signal produced by a muon traversing the station vertically).
Due to the standard
LPM effect $X_{max}$ in an EAS induced by a UHE photon is expected to occur deeper in the atmosphere 
than in the case of showers
initiated by nuclei. On the other hand, the low muon content expected in showers initiated by UHE photons
results in a lower trigger capability at larger distances, and consequently smaller $N_{stat}$ and $S_b$ when
comparing to the footprints of hadron-induced EAS. The primary type discrimination power of the above three
observables is illustrated in Fig.~\ref{fig:signatures}.
The Boosted Decision Tree (BDT) multivariate analysis method has been applied to the 
available data set
using $X_{max}$, $N_{stat}$ and $S_b$.
The BDT method has been chosen after examining several variants, as illustrated in the left panel of 
Fig.~\ref{fig:signatures-bdt}, leading to the optimal selection cut suitable for $E_\gamma>10^{18}$~eV 
(Fig.~\ref{fig:signatures-bdt}, right panel).
The few photon candidates found in this way were checked to be consistent with the proton background which led
to determining the new hybrid upper limits to photon fluxes, as seen in Fig.~\ref{fig:limits}.
The new limits are more stringent in comparison to previous ones (see Ref.~\cite{auger-photons-bleve2015}),
and the uncertainties of the limits are specified for the first time. The strong constraints on the ``top-down'' models
can be concluded under the assumptions explained in Sections~\ref{sec:intro} and \ref{sec:signatures}. For the 
first time also the photon flux predictions from one of the ``bottom-up'' scenarios (``GZK proton I'') are constrained
below $E_\gamma$=10$^{19}$~eV. The reader is referred to 
Ref.~\cite{auger-diffuse-photon-2017} for details and references.

\section{Directional blind search}
\begin{figure}[ht]
\begin{center}
\includegraphics[width=0.6\linewidth]{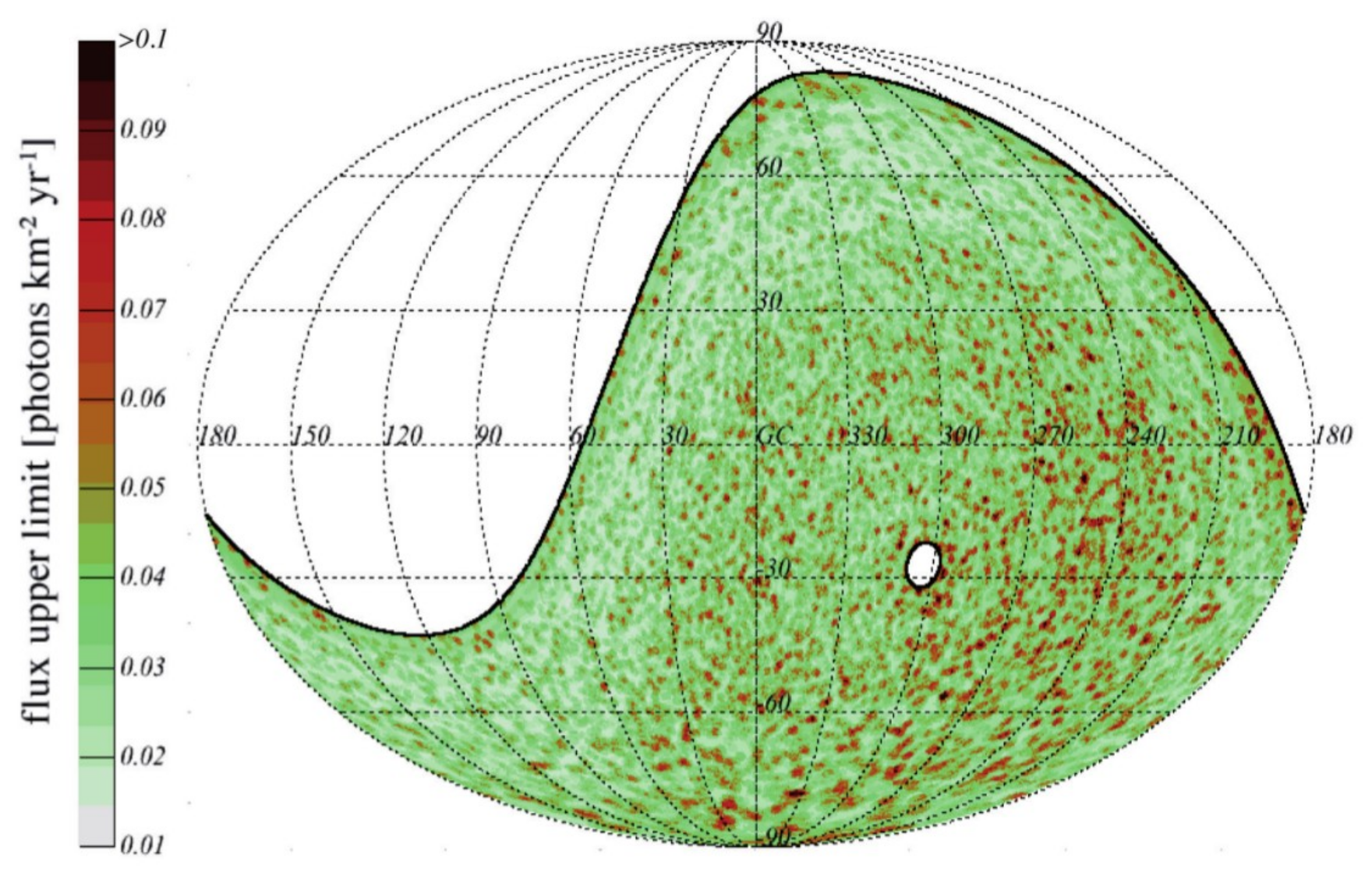}
\caption{Celestial map of photon flux upper limits in photons km$^{−2}$yr$^{−1}$ illustrated in Galactic 
coordinates.~\cite{auger-photon-pointsources-2014}}
\label{fig:directional}
\end{center}
\end{figure}

Complementing the diffuse photon flux study, the Pierre Auger Observatory performed also the blind
search for arrival directions where photon excesses could be observed \cite{auger-photon-pointsources-2014}. 
While the diffuse flux study aims at identifying all the photon candidates regardless 
their arrival directions, the directional blind search
tries to identify photon candidates grouping directionally, and thus pointing to possible photon sources. 
The sensitive search for point sources was performed within a declination
band from $-$85$^{\circ}$ to +20$^{\circ}$, and in an energy range from 10$^{17.3}$~eV to 10$^{18.5}$~eV, were
the photon fluxes are not precluded by the diffuse photon search.
No photon point source has been detected and an upper limit on the photon flux has been derived for 
every direction. In the study the method to derive a skymap of upper limits to the photon fluxes of point sources 
(Fig.~\ref{fig:directional}) was specified.

\section{Targeted search}
\begin{figure}[ht]
\begin{center}
\includegraphics[width=0.6\linewidth]{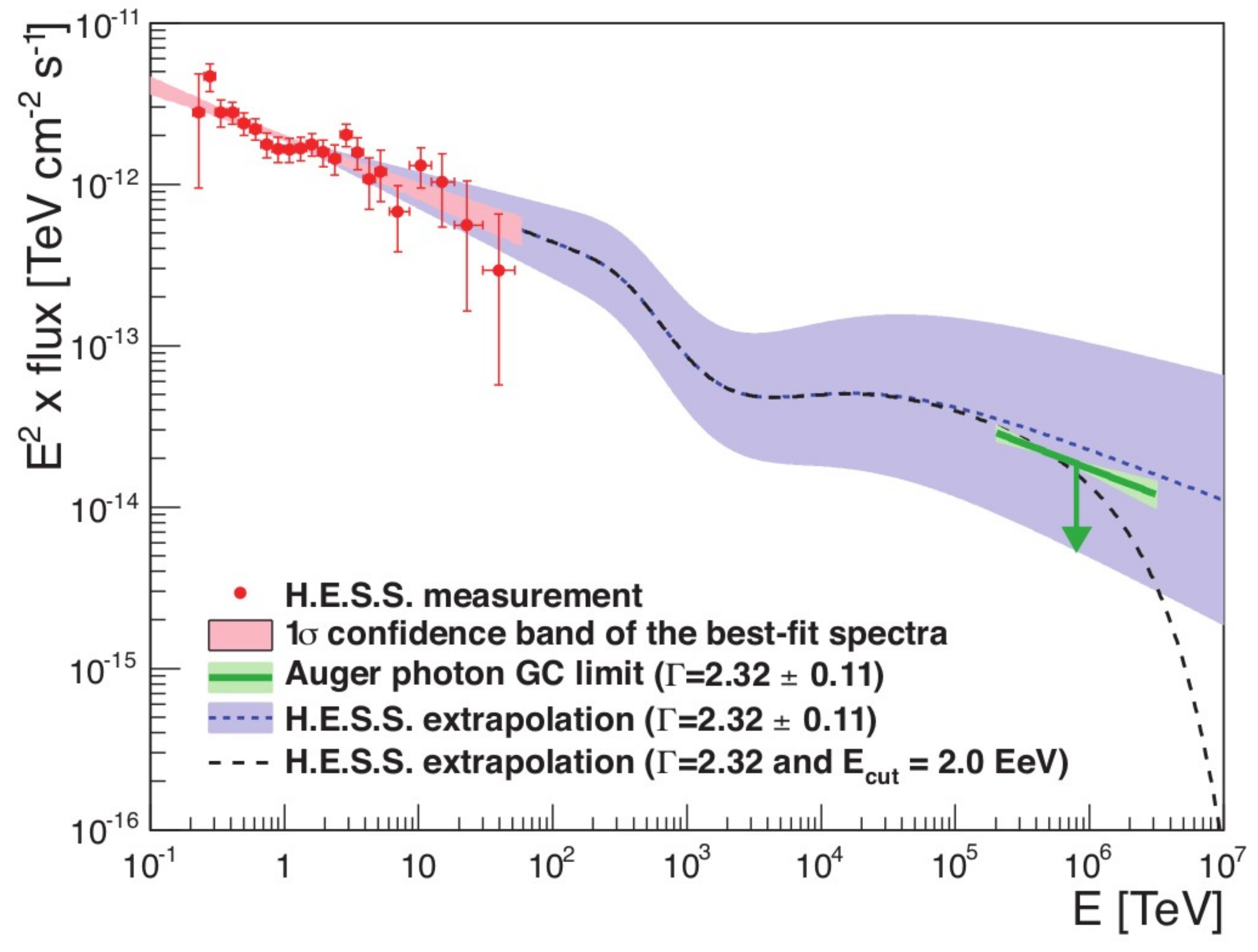}
\caption{Photon flux as a function of energy from
the Galactic center region. Measured data by H.E.S.S.
are indicated, as well as the extrapolated photon flux at
Earth in the EeV range, given the quoted spectral indices
(\cite{hess-pevatrons-nature2016}; conservatively the extrapolation
does not take into account the increase of the $p$-$p$ cross section toward higher energies). 
The Auger limit is indicated
by a green line. A variation of the assumed spectral index by
$\pm$0.11 according to systematics of the H.E.S.S. measurement
is denoted by the light green and blue band. A spectral index
with cutoff energy E$_{cut}$=2.0$\times$10$^6$~TeV is indicated as well.~\cite{auger-targeted-photon-2017}}
\label{fig:targeted}
\end{center}
\end{figure}

To reduce the statistical penalty of many trials from
that of a blind directional search in Ref.~\cite{auger-targeted-photon-2017} several Galactic and extragalactic 
candidate objects were grouped in classes and analyzed for 
a significant photon excess above the background expectation. No evidence for photon emission from candidate sources
was found and the corresponding particle and energy flux upper limits were given. 
These limits significantly constrain predictions of EeV proton emission models from
non-transient Galactic and nearby extragalactic sources, as illustrated for the particular case of the
Galactic center region (Fig~\ref{fig:targeted}).

\section{Summary and Outlook}
\label{sec:sumary}

In this report the status of the ultra-high energy photon search at the 
Pierre Auger Observatory is summarized with emphasis on the recent advances, including
the searches for the diffuse UHE photon flux, emission from discrete sources and  
targeted photon search. Although none of the searches performed so far proved an 
existence of ultra-high energy photons, the diffuse, directional and grouped upper limits to photon fluxes
provide valuable astrophysical constraints. The end of the road to discovering
UHE photons in the Pierre Auger Observatory has not yet been reached. 
The new substantial amount of data of increased sensitivity to primary mass will be acquired within the next 
several years with the AugerPrime upgraded detectors~\cite{auger-upgrade} which 
could lead either to identification of photons or
to further constraints on the ``bottom-up'' predictions. 

The UHE photon search can be continued also 
with alternative methods. UHE photon sensitivity comparable to the 
Pierre Auger Observatory is possible e.g. with the Cherenkov Telescope Array (CTA) 
\cite{uhe-photon-cta-exposure-neronov16}. 
Considering gamma ray detection technique for the UHE photon search implies 
using an alternative set of observables, unreachable by standard cosmic-ray arrays. The first studies
in this direction point to a chance for a very precise identification of a photon primary or photon pre-shower,
even event-by-event \cite{credo-gamma-rays-icrc2017,credo-general-icrc2017}. 

Another interesting scenario involving UHE photons and their interactions is based on questioning
some of the assumptions underlying the physical interpretation of the presently reported non-observation
of UHE photons. In these scenarios (e.g. LIV with rapid photon decay) UHE photons produced at astrophysical 
distances have very short lifetimes and
have a negligible chance to reach 
Earth~\cite{Klinkhamer:2008ky,chadha83-phot-decay,kostelecky2002-phot-decay,jacobson-2005-liv-rev}. 
On the other hand, the experimental verification of such models
would be possible only if there is a chance to observe at least partly the products of UHE photon interactions: 
extensive cascades of cosmic rays. Such a chance should then be determined for the scenarios leading
to UHE photon cascading in order to proceed with the experimental effort. Alternatively one could also think
of a global cosmic-ray analysis strategy dedicated to hunting for large scale time correlations
independently of the expectations from the existing theoretical models.

\section*{Acknowledgements}

PH thanks the Pierre Auger Collaboration, in particular Daniel Kuempel and Marcus Niechciol for help 
in summarizing the status of the UHE photon search, and Mikhail V. Medvedev, \L{}ukasz Bratek, and David
d'Enterria for inspiring discussions about the perspectives of further studies in this direction.


\end{document}